\documentclass{pasj00}

\DeclareAbbreviation\ibvs{Inf. Bull. Variable Stars}
\def\ASPConf#1#2{ASP Conf. Ser. #1, #2}
\def\PublisherASP{San Francisco: ASP}
\def\PublisherReidel{Dordrecht: D. Reidel Publishing Company}

\begin{document}
\SetRunningHead{T. Kato et al.}{Late Superhumps in WZ Sge-Type Dwarf Novae}

\Received{200X/XX/XX}
\Accepted{200X/XX/XX}

\title{Late Superhumps in WZ Sge-Type Dwarf Novae}

\author{Taichi \textsc{Kato}}
\affil{Department of Astronomy, Kyoto University,
       Sakyo-ku, Kyoto 606-8502}
\email{tkato@kusastro.kyoto-u.ac.jp}

\author{Hiroyuki \textsc{Maehara}}
\affil{Kwasan and Hida Observatories, Kyoto University, Yamashina,
       Kyoto 607-8471}

\email{\rm{and}}

\author{Berto \textsc{Monard}}
\affil{Bronberg Observatory, PO Box 11426, Tiegerpoort 0056,
       South Africa}



\maketitle

\begin{abstract}
   We report on the detection of very stable modulations with periods
unexpectedly ($\sim$0.5 \%) longer than superhump periods during the
slowly fading stage of WZ Sge-type superoutbursts in three systems,
GW Lib, V455 And and WZ Sge.  These periods are naturally explained
by assuming that these modulations are superhumps arising from matter
near the tidal truncation radius.  This finding provides
an additional support to the hypothetical idea of expansion of the
accretion disk well beyond the 3:1 orbital resonance in some low
mass-ratio systems.  Combined with the effect of 2:1 resonance,
we present an explanation of the origin of positive period derivatives
in certain short-period SU UMa-type dwarf novae. 
\end{abstract}

\section{Introduction}
   Dwarf novae (DNe) are a class of cataclysmic variables (CVs), which are
close binary systems consisting of a white dwarf and a red-dwarf secondary
transferring matter via the Roche-lobe overflow.
SU UMa-type dwarf novae are a class of DNe, which show superhumps during
their long, bright outbursts (superoutbursts)
[see e.g. \citet{vog80suumastars}; \citet{war85suuma} for basic observational
properties].  The origin of superoutbursts and superhumps in SU UMa-type
dwarf novae is basically understood as a consequence of thermal and tidal
instabilities in the accretion disk (\cite{osa89suuma}; \cite{osa96review}),
the latter being excited by the 3:1 orbital resonance in the disk
(\cite{whi88tidal}; \cite{hir90SHexcess}; \cite{lub91SHa}).
The basic observational properties of ordinary SU UMa-type dwarf novae
have well been reproduced by this picture.

   WZ Sge-type dwarf novae (see e.g. \cite{bai79wzsge}; \cite{dow90wxcet};
\cite{kat01hvvir}) are a subgroup of dwarf novae characterized by
large-amplitude (typically $\sim$ 8 mag) superoutbursts with very long
(typically $\sim$ 10 yr) recurrence times.  Although WZ Sge-type
dwarf novae are recognized as a subgroup of SU UMa-type dwarf novae,
WZ Sge-type dwarf novae are known to have a number of properties hardly,
but not necessarily exclusively, observed in ``textbook'' SU UMa-type
dwarf novae.  These properties include: (1) the presence of early
superhumps \citep{kat02wzsgeESH}, (2) (sometimes repetitive) rebrightenings
\citep{kat04egcnc}, (3) nearly constant to positive period derivative
($P_{\rm dot} = \dot{P}/P$) of superhumps (\cite{kat01hvvir};
\cite{kat03v877arakktelpucma}), and (4) long-lasting fading tails.

   The implication of phenomenological relations between some of these
properties was first addressed by \citet{kat98super} [see also
\citet{kat04egcnc} for more discussions].  \citet{kat98super} presented
an idea that the accretion disk can expand
beyond the 3:1 resonance during energetic outbursts in low mass ratio
($q = M_2/M_1$) systems exemplified by WZ Sge-type dwarf novae. 
They argued that the matter beyond the 3:1 resonance can serve as
a reservoir supplying matter to the inner disk resulting rebrightenings,
and the eccentricity wave propagating outward the 3:1 resonance can
explain positive $P_{\rm dot}$ of superhumps.  This idea thus has
a possibility to naturally explain many of peculiar properties of
WZ Sge-type dwarf novae.  \citet{hel01eruma} further introduced an idea
of decoupling of thermal and tidal instabilities beyond the 3:1 resonance,
and extended the application to ER UMa-type dwarf novae, another subclass
of SU UMa-type dwarf novae with similarly low $q$.

   In line with these ideas, \citet{osa02wzsgehump} and
\citet{osa03DNoutburst} presented
an overview of an WZ Sge-type outburst based on the disk-instability
model of SU UMa-type dwarf novae (\cite{osa89suuma}; \cite{osa95wzsge}),
and proposed a new conceptual scheme of classifying SU UMa-type
dwarf novae based on $q$ and achievable disk radius.
\citet{osa03DNoutburst} also explored the dependence of outburst
properties on the matter reaching beyond the 3:1 resonance, and presented
a scheme of understanding a variety of superoutbursts.
The expanded disk beyond the 3:1 resonance in WZ Sge-type dwarf novae
and related objects has thus been favored by theoretical sides.
Observational evidence, however, for such an expanded disk had long
been rather scarce [see \citet{kat04egcnc} for description of historical
observations],
while the recent discovery of different $P_{\rm dot}$ of
superhumps in a variety of superoutburst in the same system
\citep{uem05tvcrv} well matched the scenario by \citet{osa03DNoutburst},
thus strengthening the expanded disk beyond the 3:1 resonance.
The infrared excess during the late stage of WZ Sge-type outbursts
(e.g. \cite{uem08j1021}; \cite{uem08alcom}) also supports this idea.

   In 2007, two spectacular superoutbursts of WZ Sge-type dwarf novae,
namely GW Lib and V455 And occurred (\cite{waa07gwlibiauc};
vsnet-alert 9530).  During the later course of these
outbursts, we discovered ``late superhumps'' with unexpectedly long
periods and with exceptionally high coherence and stability in their
periods.  In this letter, we present an interpretation of these
late superhumps originating from the very matter beyond the 3:1 resonance.

\section{Late Superhumps in GW Lib and V455 And}

\begin{figure}
  \begin{center}
    \FigureFile(88mm,60mm){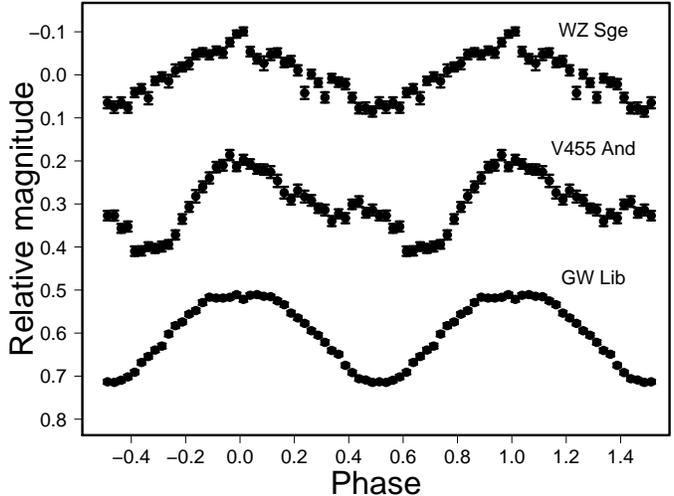}
  \end{center}
  \caption{Phase-averaged profiles of late superhumps.
  }
  \label{fig:lshave}
\end{figure}

   Both outbursts of GW Lib and V455 And were extensively observed
photometrically by the VSNET Collaboration using a network of small
telescopes \citep{VSNET}.  The details of these observations will be
presented in separate papers.  In addition to early superhumps
signifying the WZ Sge-type dwarf nova (in V455 And), ordinary superhumps
during the superoutburst plateau (in GW Lib and V455 And), we detected
additional hump features persisting after the termination of the
superoutburst plateau.  The periods of the last modulations were
substantially longer than those of orbital periods, and even longer
than those of ordinary superhumps, ruling out the possibility of
modulations arising from orbital humps.
After detecting unequivocal signals in GW Lib and V455 And, we reanalyzed
the archival data of the 2001 outburst of WZ Sge (cf. \cite{ish02wzsgeletter})
and detected a corresponding signal.\footnote{
  \citet{pat02wzsge}, using a slightly different set of data from ours,
  also detected the same periodicity and gave a period of 0.05736(5) d
  but they identified it as being traditional late superhumps
  (\cite{vog83lateSH}; \cite{hes92lateSH})
}
The representative periods of these late superhumps are summarized
in table \ref{tab:lsh}.
Figure \ref{fig:lshave} shows a comparison of averaged profiles
of late superhumps in these three WZ Sge-type dwarf novae.

\begin{table*}
\caption{Periods of late superhumps.}\label{tab:lsh}
\begin{center}
\begin{tabular}{ccccccc}
\hline\hline
Object & $P_{\rm orb}$$^*$ & $P_{\rm sh}$$^\dagger$ & $P_{\rm lsh}$$^\ddagger$
       & JD range$^\S$
       & $q$ (adopted) & $R_{\rm d}$$^\|$ \\
\hline
GW Lib   & 0.05332(2) & 0.053925(4) & 0.054156(1) & 2454230--245 & 0.062 & 0.58 \\
V455 And & 0.05630921(1) & 0.05697(1) & 0.057280(4) & 2454367--378 & 0.064 & 0.61 \\
WZ Sge   & 0.0566878460(3) & 0.05721(5) & 0.057408(4) & 2452182--210 & 0.050 & 0.59 \\
\hline
 \multicolumn{7}{l}{$^*$ Orbital period (d).} \\
 \multicolumn{7}{l}{$^\dagger$ Period of ordinary superhumps (d) (see text).} \\
 \multicolumn{7}{l}{$^\ddagger$ Period of late superhumps (d).} \\
 \multicolumn{7}{l}{$^\S$ Range of JD for determining the periods of
 late superhumps.} \\
 \multicolumn{7}{l}{$^\|$ Radius (unit in binary separation) corresponding
 to the fractional period excess of} \\
 \multicolumn{7}{l}{late superhumps (see text).} \\
\end{tabular}
\end{center}
{\bf References:}
  GW Lib: \citet{tho02gwlibv844herdiuma}; vsnet-outburst 7773;
  V455 And: \citet{ara05v455and}; vsnet-alert 9326, 9642;
  WZ Sge: \citet{ste01wzsgesecondary}; \citet{pat02wzsge};
          \citet{ish02wzsgeletter}
\end{table*}

\begin{figure}
  \begin{center}
    \FigureFile(88mm,60mm){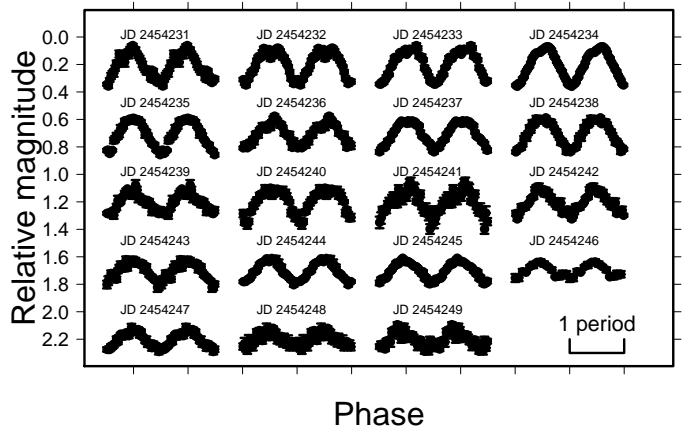}
  \end{center}
  \caption{Example of nightly profiles of late superhumps in GW Lib.
  Note the stability of phases.
  }
  \label{fig:gwlib}
\end{figure}

   Figure \ref{fig:gwlib} shows nightly profiles of late superhumps in
GW Lib.  Although the figure covers 18 d ($\sim$330 cycles),
the stability of the hump phases is striking.  The period changes were
almost absent, or the variation was even more exactly periodic
than ordinary superhumps.
The stability of the periods was less convincing in higher inclination
systems V455 And and WZ Sge due to the strong overlapping orbital
signals, though the $O-C$ diagram shown in \citet{pat02wzsge} suggests
a similar degree of stability.

\section{Interpretation of WZ Sge-Type Late Superhumps}

   The most striking feature of late superhumps common to all three
WZ Sge-type dwarf novae is their long periods.  The fractional superhump
excesses ($\epsilon=P_{\rm SH}/P_{\rm orb}-1$) lie between 1.2\% and 1.7\%,
typically $\sim$0.5\% longer than the ordinary superhump period
of the respective object.
These very long periods are hard to understand, since period excesses
are generally understood to arise from a precessing eccentric disk
(\cite{osa85SHexcess}; \cite{hir93SHperiod}),
whose angular velocity ($\omega_{\rm p}$) has radial dependence of
$\omega_{\rm p} \propto R_{\rm d}^{3/2}$, where $R_{\rm d}$ is
the radius of the accretion disk \citep{osa85SHexcess}, and the disk
is expected to have shrunken after the outburst\footnote{
  \citet{how96alcom} suggested an interpretation assuming the expansion
  of the disk beyond the radius of 3:1 resonance in a different context
  of disk evolution.
}.

   The situation, however, drastically varies by simply assuming the
accretion disk beyond the 3:1 resonance contributes to these late
superhumps.  Using the radial dependence of $\omega_{\rm p}$
and assuming that ordinary superhumps
arise from the radius of 3:1 resonance ($R_{3:1}$),
we can estimate the radii
where these long-period late superhump originate.\footnote{
  \citet{mur00SHprecession} has pointed out that the simple analytical
  precession rate as in \citet{osa85SHexcess} is inadequate.
  The radial dependence given in
  \citet{mur00SHprecession} gives slightly
  (10--20\% expressed in $R_{\rm d}/R_{3:1}$) smaller radii
  than those listed in table \ref{tab:lsh}.  \citet{smi07SH} reported
  from numerical simulations that observed superhump period reflects
  a radius slightly smaller than that of $R_{3:1}$, which
  difference could bring a similar order of uncertainty.
  For simplicity, we simply used the radial dependence in
  \citet{osa85SHexcess} and $R_{3:1}$ for rescaling the observed periods
  into radii as the first approximation since $q$ and other parameters
  have comparable uncertainties.  
}
In these calculations, we adopted mean period of ordinary superhumps
(in table \ref{tab:lsh}) near the onset of the superoutburst plateau,
when superhump period is expected to reflect the precession rate at
a radius close to $R_{3:1}$ \citep{uem05tvcrv}.  We used the typically
adopted $q$ = 0.050(15) for WZ Sge (cf. \cite{pat05SH}) and
rescaled $\epsilon$ of other objects into $q$ using the relation in
\citet{pat05SH}.
The following discussion is not very sensitive to these assumptions.

\begin{figure}
  \begin{center}
    \FigureFile(88mm,88mm){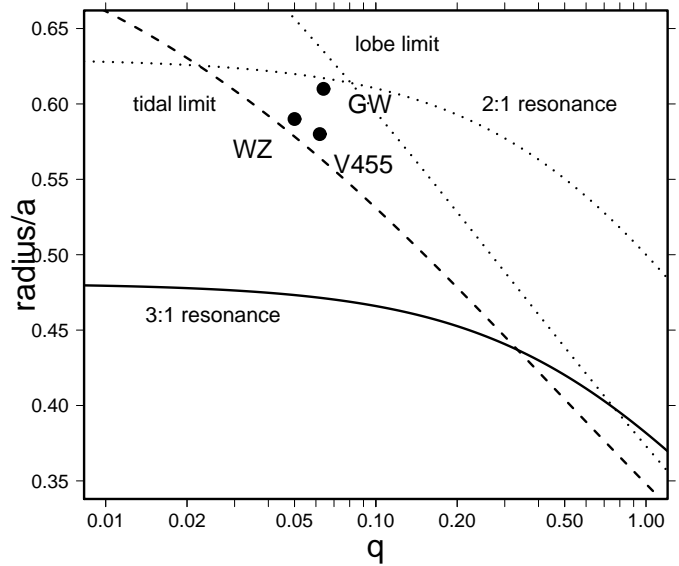}
  \end{center}
  \caption{Disk radii corresponding to periods of late superhumps
  in WZ Sge-type dwarf novae.
  The lines of 3:1 and 2:1 resonances,
  and the tidal limit are the same as in \citet{osa02wzsgehump}.
  The lobe limit represents the minimum radius of the Roche lobe
  on the orbital plane.  The estimated disk radii of three WZ Sge-type
  dwarf novae are located close to the tidal limit.
  }
  \label{fig:lshr}
\end{figure}

   The estimated radii are shown in figure \ref{fig:lshr}.
The estimated disk radii of all three WZ Sge-type dwarf novae are
located close to the tidal limit.  The result strongly suggests
that these late superhumps are better understood to arise from
at large radii rather than classical late superhumps reflecting
the luminosity `hot spot' as in \citet{vog83lateSH}.  The high
stability of the periods can be naturally understood as the result
of outer disk radius fixed at the tidal limit.
As discussed in \citet{osa03DNoutburst},
tidal truncation simply inhibits the disk to expand beyond
the corresponding radius but does not cause the disk to contract below
this limit.  This action provides a favorable condition to
maintain the constant precession rate (i.e. period) for a long time.

\section{Positive Period Derivatives of Superhumps -- Toward Unified
         Picture of Variety of Superoutbursts}

   Some of WZ Sge-type dwarf novae and short-period SU UMa-type
dwarf novae show positive $P_{\rm dot}$ of ordinary superhumps
(\cite{kat01hvvir}; \cite{kat03v877arakktelpucma}).
\citet{uem05tvcrv} interpreted that this period increase reflects
the propagation of eccentricity wave beyond the $R_{3:1}$
[see also \citet{kat98super}].  It was unclear, however, what
regulates the wave propagation.

   We here propose that the 2:1 resonance occurring in the outermost
disk governs the range where 3:1 resonance can appear.
Following the discussion in \citet{osa03DNoutburst}, the 2:1 resonance
suppresses the growth of 3:1 resonance.  In extreme cases as in the
early stage of WZ Sge-type superoutbursts, the 2:1 resonance can be
strong enough to entirely suppress the 3:1 resonance.  As the
superoutburst proceeds and the matter is swept from the outer region,
this effect weakens and enables the 3:1 resonance to grow in the inner
region.  As a natural extension to this interpretation, we consider
that the strength of the 2:1 resonance determines the limit of radius
inside which eccentricity can grow.  When this radius is smaller
than $R_{3:1}$, ordinary superhumps cannot grow
[this corresponds to the delay in appearance of ordinary superhumps
in WZ Sge-type dwarf novae \citep{osa03DNoutburst}].

   Ordinary superhumps grow when this limit becomes larger than the
radius of the 3:1 resonance.  Subsequent behavior will depend on
the mass and state of the disk beyond $R_{3:1}$.
In extreme cases as in typical WZ Sge-type
dwarf novae, the strength of the 2:1 resonance can be strong enough
to accrete much of the matter beyond $R_{3:1}$ (as in the snow-plowing
effect in SU UMa-type superoutburst, \cite{osa89suuma}), leading to
a cold, low-mass disk outside $R_{3:1}$.  Even if the eccentricity
wave can propagate into this region, it would not produce a strong superhump
signal.  This could explain why $P_{\rm dot}$ is almost zero in most
WZ Sge-type dwarf novae.  In less extreme cases with similar or slightly
larger $q$ but with weaker 2:1 resonance (assuming that the disk can
transiently expand beyond the tidal truncation radius, see also
\cite{osa03DNoutburst}), the resonance itself or its effect
diminishes before the matter beyond $R_{3:1}$ is efficiently
cleared.  In such cases, still ionized sufficient matter beyond
the 3:1 resonance could produce ordinary superhumps with increasing
periods as the limit moves outward.  The recent discovery of short-lived
early superhumps and positive $P_{\rm dot}$ in a larger $q$-system, BC UMa
\citep{mae07bcuma}, as well as an earlier example of RZ Leo
\citep{ish01rzleo}, supports this idea.  In most SU UMa-type dwarf novae
with larger $q$, the mass beyond $R_{3:1}$ is too small to
show increasing periods regardless of the condition of the 2:1 resonance.
This interpretation can explain why systems with large positive
$P_{\rm dot}$ are restricted to a relatively small region with intermediate
orbital periods.

   This interpretation, as in \citet{osa03DNoutburst}, proposes that
suppression of the 3:1 resonance by the 2:1 resonance is the significant
cause of long delays of appearance of ordinary superhumps in WZ Sge-type
superoutburst, rather than the effect of slow growth of the 3:1 resonance
in small $q$ systems in traditional view \citep{osa96review}.
In systems that somehow enables 2:1 resonance,
this interpretation predicts variable delay times of appearance of
ordinary superhumps even in the same object depending on the disk mass
at the onset of superoutbursts (\cite{osa03DNoutburst}; \cite{uem05tvcrv}).
If the growth rate of the 3:1 resonance depending $q$ is the main
factor determining this delay, such variation would not be expected.
This prediction seems to be confirmed by past observation in SW UMa
($q$ = 0.11, \cite{pat05SH}): the delay was 6--11 d for a very bright
($m_{\rm v} =$ 9.6 at maximum) superoutburst in 1986 \citep{rob87swumaQPO},
5--6 d for a bright ($m_{\rm v} =$ 10.1) one in 2006 September
(vsnet-alert 9018) while it was only $\leq$3 d for the 2002 one
(\cite{kat92swumasuperQPO}, $m_{\rm v} =$ 10.8 at maximum).
Systematic future observations of superoutbursts of various scales
will be a key in testing this interpretation.

\vskip 3mm

   The author is grateful to observers of VSNET Collaboration who
supplied vital data.

\end{document}